\documentstyle[epsf,epsfig,aps,prb,multicol]{revtex}
\newcommand{\bea}{\begin{eqnarray}}
\newcommand{\eea}{\end{eqnarray}}

\begin{document}

\title{Tomonaga-Luttinger liquid with reservoirs in a multi-terminal geometry.}

\author{K-V Pham, F. Pi\'echon, K-I Imura, P. Lederer.}

\address{Laboratoire de Physique des Solides, B\^at. 510, CNRS, 
universit\'e Paris-sud, 91405 Orsay, France}

\maketitle

\begin{abstract}
We propose a formalism which uses boundary conditions imposed on the
Luttinger liquid (LL) to describe the transport properties of a LL
coupled to reservoirs. The various boundary conditions completely
determine linear transport in the joint system reservoirs+LL. As an
illustration we consider an exactly solvable microscopic model in
a multi-terminal geometry for which such boundary conditions can be
explicitly derived; in this model the Landauer-B\"uttiker formalism
fails: if it were valid, the relation between the conductance matrix elements and the reflection and transmission coefficients could yield 
negative probabilities. We then apply our 
formalism to a discussion of shot noise through an
impurity in a LL connected to two reservoirs.
\end{abstract}


\begin{multicols}{2}

\section{Introduction}

For Fermi liquids coupled to many reservoir leads 
linear transport properties 
can be described by the Landauer-B\"uttiker formalism whose
key idea is to relate transport properties of each single electron
at the Fermi level to some transmission and reflection probabilities
that characterize the scattering properties of electrons \cite{1,2}.
The Landauer-B\"uttiker formalism states that the conductance matrix
elements \( G_{np} \) defined for each conducting channel 
by \( i_{n}=\sum_{p}G_{np}V_{p} \)
(where \( i_{n} \) and \( V_{p} \) are respectively the current
injected by reservoir \( n \) and the voltage of reservoir \( p \))
are related to the transmission ($T_{np}$) and reflection ($R_n$) 
probabilities of single electron:
\bea
G_{nn} & =\frac{e^{2}}{h} & \left( 1-R_{n}\right) \label{1} \\
G_{np} & =-\frac{e^{2}}{h} & T_{np}. \label{2} 
\eea
Therefore: (i) \( G_{nn}>0 \) and \( G_{np}<0 \) and both are bounded
by \( e^{2}/h \); (ii) in the absence of any backscattering \( G_{nn} \)
has a finite universal value \(G_0=e^{2}/h \) which leads in a two-terminal
geometry to the so-called universal contact resistance that reflects
the (inelastic) relaxation process of each electron that leaves the
sample and enters a given reservoir.

For strongly correlated electronic systems such as the Luttinger Liquid
(LL) the single particle scattering matrix approach essential to the
Landauer-B\"uttiker formalism becomes inadequate because in a LL the
single-particle Green function has no quasiparticle pole which means
that electrons do not propagate \cite{3}.

Nevertheless, the Landauer-B\"uttiker formalism has been widely used
to interpret recent two and multi-terminal linear transport measurements
\cite{key-15,key-19,key-20}
on \emph{physical systems that are believed to be good examples of
LL} \cite{key-16,key-23}. More precisely several experiments on quantum
wires fabricated by cleaved edge overgrowth found two-terminal conductance
plateaux at nonuniversal values: \( G=0.8 \), \( 0.85 \), \( 0.9 \)
 \cite{key-19} and as low as \( 0.45 \) in units of the quantum of
conductance \( G_{0} \) (per channel) \cite{key-16}. In
Ref.\cite{key-19} results for a three terminal geometry are also
reported and show that \( R_{n}\neq 0 \) 
if interpreted in the Landauer-B\"uttiker
frame. Assuming then that electrons in the wire are 
non-interacting these values
can be understood by invoking some backscattering due to the coupling
of the ballistic wire to the two-dimensional reservoirs \cite{key-19}.
Similarly individual single wall carbon nanotubes were found with
either a two-terminal conductance \( G=G_{0} \) \cite{key-20} but
also with \( G\sim 0.5-0.6G_{0} \) \cite{key-21}.

If one believes that quantum wires and carbon nanotubes constitute
realizations of the LL the interpretation of these non universal ohmic
conductance plateaux poses a serious problem in two respects: (i)
if the smaller conductances are interpreted as resulting from some
backscattering (in agreement with a Landauer-B\"uttiker interpretation)
then this is contradictory to the expectation that in a LL backscattering
should lead to power law (non-ohmic) corrections; (ii) moreover in
the framework of the LL theory several earlier theoretical papers
found that the two-terminal conductance of a LL should be unaffected
by the interactions and stick to a universal value \( G_{0} \) yielding
the same contact resistance as in non-interacting systems \cite{key-14}. 

Actually a theoretical analysis of the transport
properties of a LL in a multi-terminal geometry 
has not really been developped. 

In this paper we develop a formalism to describe the linear transport
properties of a LL coupled to an arbitrary number of electrodes, which
play the role of charge reservoirs for the LL. Instead of a full-fledged
microscopic approach which is obviously unassailable we propose a
formalism in which the coupling of a LL to reservoirs is taken into
account by considering a LL subjected to boundary conditions. Our
approach generalizes several earlier papers which take a similar stand
for the modelling of a coupled system \cite{key-8,key-9,key-10,key-11,key-12}.

Our basic idea is as follows: (i) we observe that in a LL a chiral
decoupling occurs, i.e. each chirality is independent of the other
and is responsible for the transport of current in one direction,
and therefore the transport properties are completely specified by
the associated chiral chemical potentials. (ii) Imposing linear relations
to the terminal voltages determines the values of these chiral potentials:
each set of relation can therefore be seen as a set of boundary conditions
imposed on the LL. We consider that each set of boundary conditions
characterizes some kind of coupling of the electrodes to the LL. In
other words \emph{as far as the transport properties are concerned}
we view a given set of boundary conditions as corresponding to a universality
class for the total system consisting of both the LL and the electrodes.
That formalism is explained in \textbf{section 2}.

To substantiate our views we consider in \textbf{section 3} an exactly
solvable model of strongly correlated fermions coupled to an arbitrary
number of terminals and \emph{for which such boundary conditions can
be derived explicitly}. We find that in the toy model of \textbf{section
3} the signs of the conductance matrix elements are not fixed: if
we were to define reflection and transmission coefficients \( R_{n} \)
and \( T_{np} \) through equations (\ref{1},\ref{2}) we would get
negative probabilities. That the probabilistic view of transport given
by the Landauer-B\"uttiker formalism fails is hardly surprising: transport
in a LL is not ensured by electrons (or Landau quasiparticles) but
by fractional excitations akin to those of the Fractional Quantum
Hall Effect (FQHE)\cite{4}.

In \textbf{section 4} we specialize to the two-terminal geometry.
We show that the boundary conditions are equivalent in that geometry
to specifying contact resistances at the source and the drain. We
explain how our formalism allows us to classify earlier theories: theories
yield distinct conductances because they belong to different universality
classes and correspond to distinct boundary conditions. We then consider
a LL with an impurity which provides some backscattering within the
system but not at the two contacts. We then discuss implications of
our formalism in \textbf{section 5} for the various theories of shot
noise through a single impurity in a LL.

\section{Boundary conditions in a multi-terminal geometry.}

\subsection{Chiral chemical potentials for the LL.}

We consider the standard LL hamiltonian density:
\begin{equation}
\label{hb}
H=\hbar \frac{u}{2}\left[ K^{-1}\left( \partial _{x}\Phi \right) ^{2}+K\Pi ^{2}\right] 
\end{equation}
where we have introduced the standard phase field \( \Phi  \) related
to the electron density by: \( \delta \rho (x,t)=\partial _{x}\Phi /\sqrt{\pi } \)
, and its conjugate canonical momentum \( \Pi  \). We also introduce
the chiral densities: \begin{equation}
\label{rho}
\rho _{\pm }=\frac{1}{2\sqrt{\pi }}\left( \partial _{x}\Phi \pm K\Pi \right) .
\end{equation}
It is easy to check that these densities are indeed chiral and to
show that they are related to the densities of electrons at the right
and left Fermi points by: \begin{equation}
\label{rho2}
\left( \begin{array}{c}
\rho _{+}\\
\rho _{-}
\end{array}\right) =Q\left[ K\right] \left( \begin{array}{c}
\rho ^{0}_{+}\\
\rho ^{0}_{-}
\end{array}\right) ,
\end{equation}
where: \begin{equation}
\label{q}
Q[a]=\frac{1}{2}\left( \begin{array}{cc}
1+a & 1-a\\
1-a & 1+a
\end{array}\right) .
\end{equation}
In terms of these chiral variables the LL hamiltonian is completely
decoupled into two chiral hamiltonians. Current will then be induced
in the LL by adding conjugate variables to \( H \):
\begin{equation}
\label{dh}
\delta H=-\mu _{+}\rho _{+}-\mu _{-}\rho _{-}
\end{equation}
which defines chiral chemical potentials; as usual when defining chemical
potentials they correspond to the energy needed for the creation of
a unit particle of a given chirality \emph{within the LL} (\( Q_{\pm }=\int dx\: \rho _{\pm }=1) \)
. They are not defined as the reservoir chemical potentials. Minimization
of the hamiltonian leads easily to the relation:\begin{equation}
\label{courant}
I=\frac{Ke}{h}\left( \mu _{+}-\mu _{-}\right) .
\end{equation}
Therefore the two-terminal conductance of a LL as measured against
these chiral chemical potentials is:\begin{equation}
G_{K}=\frac{Ie}{\mu _{+}-\mu _{-}}=K\frac{e^{2}}{h}.
\end{equation}
Ii is crucial to realize that the experimentally measured conductance
\( G_{2t}=\frac{Ie}{\mu _{S}-\mu _{D}} \) and \( G_{K} \)
need not be identical. The chiral chemical potentials (associated
with the quasiparticles driving the current in the LL) will indeed
usually differ from the reservoir chemical potentials. 

\subsection{Boundary conditions.}

We now consider \( N \) terminals at respective voltages \( V_{n} \).
The geometry is indifferent and is either that of a loop or that of
a rod (see Figures 1 and 2). Due to current injection at each terminal
the chiral chemical potentials may change their values across them:
accordingly we add indices so that \( \mu _{n}^{+} \) and \( \mu _{n}^{-} \)
are on the left of terminal \( n \) and change to \( \mu _{n+1}^{+} \)
and \( \mu _{n+1}^{-} \) on the right of the same terminal. Depending
on the geometry for \( N \) terminals there will \( 2N \) or \( 2N-2 \)
chiral chemical potentials.
\begin{eqnarray}
\mu ^{+}_{n} & = & \sum_p a^{p}_{n}\: eV_{p} \label{bc1} \\
\mu ^{-}_{n} & = & \sum_p b^{p}_{n}\: eV_{p}
\end{eqnarray}
The Fermi energy of the LL is taken as the origin of the voltages. These equations simply express
an equilibrium condition for the chemical potentials of the LL with
those of the reservoirs. Our point of view is therefore in a sense
a thermodynamical one.

The joint system (LL+electrodes) is therefore reduced to an isolated
LL (i.e. with no electrodes) on which these boundary condtions are
imposed. The advantage of such an approach is that it avoids the complications
of microscopics. To borrow the language of the renormalization group
we view the boundary conditions as characterizing universality classes
of the system (LL+electrodes). Any set of boundary conditions labels
a fixed point of the joint system and completely determines linear
transport as will be shown below.

Since \( eV_{p} \) is the chemical potential of reservoir \( p \)
charge conservation results trivially in:
\begin{equation}
1=\sum _{p}a^{p}_{n}=\sum _{p}b^{p}_{n}.
\label{charge}
\end{equation} 
The current injected at each terminal is simply the difference between
the current circulating on the right of the terminal and the current
on the left of it:
\begin{equation}
i_{n}=K\frac{e}{h}\left[ \left( \mu _{n+1}^{+}-\mu _{n+1}^{-}\right) -\left( \mu _{n}^{+}-\mu ^{-}_{n}\right) \right] .
\label{i2}
\end{equation}
Taking into account the boundary conditions eq.(\ref{bc1}) leads
to:
\begin{eqnarray}
G_{nn}= & K\frac{e^{2}}{h}\left( a^{n}_{n+1}-a^{n}_{n}+b^{n}_{n}-b^{n}_{n+1}\right)  & \label{g} \\
G_{np}= & K\frac{e^{2}}{h}\left( a^{p}_{n+1}-a^{p}_{n}+b^{p}_{n}-b^{p}_{n+1}\right) .
\end{eqnarray}
It is trivial to check that the following identities are obeyed:
\begin{eqnarray}
\sum _{p}G_{np} & = & 0\label{g2} \\
\sum _{p}G_{pn} & = & 0.
\end{eqnarray}
The first identity implements the fact that the origin of potentials
is arbitrary (gauge invariance); the second identity is just current conservation.

The requirement that Onsager-Casimir relations and positive semidefiniteness
of the conductance matrix (this ensures dissipation of energy) 
be realized evidently puts further constraints
on the coefficients \( a_{n}^{p} \) and \( b_{n}^{p} \) (such as
positivity of the diagonal elements \( G_{nn} \)) but we will not
use them in this paper.

These boundary conditions can then be straightforwardly generalized
to the case of several conducting channels as in carbon nanotubes.

\section{A multi-terminal model. }

We introduce in this part a model derived from one initially proposed
by Chamon and Fradkin for the FQHE \cite{key-34}. The model consists
in a chiral Luttinger liquid in contact with N reservoirs through
point contacts in the strong coupling limit (no backscattering at
the contact) (see Figure 3). We will show that our boundary conditions
formalism can be explicitly derived for this model of strongly correlated
electrons; moreover we show that the conductance matrix elements can
surprisingly have signs forbidden in the Landauer-B\"uttiker formalism. 

\subsection{Some useful preliminaries.}

We review first relevant results for the lagrangian of a chiral LL
coupled to a single reservoir \cite{key-34}:
\begin{eqnarray}
L & = & L_{edge}+L_{reservoir}+L_{tunnel}\label{l} \\
L_{edge} & = & \frac{1}{4\pi }\partial _{x}\varphi \left( \partial _{t}-u\partial _{x}\varphi \right) \\
L_{tunnel} & = & \Gamma \delta (x)e^{i\frac{e\Delta Vt}{\hbar }}\Psi _{edge}^{+}(x,t)\Psi _{reservoir}+h.c.
\end{eqnarray}
The lagrangian for the reservoir is taken to be that of free chiral
electrons: the electrons are chiral because free electrons on a half-line
are tantamount to free chiral electrons on a full line. 
\( \Delta V=V_{reservoir}-V_{edge}^{in} \) 
is the potential difference between the reservoir and the incoming
edge electrons. The edge electron operator is:
$\Psi _{edge}=e^{-i\frac{1}{\sqrt{\nu }}\varphi }$
where \( \nu =\frac{1}{2n+1} \)is the Quantum Hall state filling
factor.

In the strong coupling limit, the tunneling current can then be shown
to be:\begin{equation}
\label{i_tunnel}
i_{tunnel}=\frac{2\nu }{\nu +1}\frac{e^{2}}{h}\left( V_{reservoir}-V^{in}_{edge}\right) 
\end{equation}
which can also be expressed in terms of the chemical potential \( V^{out}_{edge} \) after the tunneling event:
\begin{equation}
\label{i_tunnel2}
i_{tunnel}=\nu \frac{e^{2}}{h}\left( V_{edge}^{out}-V_{edge}^{in}\right) .
\end{equation}
That expression for the tunneling current is valid at the strong coupling
fixed point (\( \Gamma =\infty  \)): therefore transmission is always
perfect with no backscattering at the contact. Away from the fixed
point (at finite \( \Gamma  \)), there would be a contribution due
to backscattering which vanishes as \( \frac{1}{\Gamma ^{K}}\Delta V^{2(K-1)} \). \emph{In our model we will also work in this ohmic limit of no backscattering}.

Chamon and Fradkin then consider \( N_{R} \) such tunnel point contacts,
each at the same potential \( V_{reservoir}=V_{D} \) and then \( N_{L} \)
similar point contacts at the potential \( V_{reservoir}=V_{S} \).
In this way, they modelize a quantum Hall bar at filling \( \nu  \)
with two 2D reservoirs on the left and at the right of the sample,
which are connected to the Hall bar through respectively \( N_{L} \)
and \( N_{R} \) point contacts. An assumption underlying this model
is therefore that the point contacts are incoherent, i.e. each tunneling
at each point contact can be considered independently of the other
point contacts. The two-terminal conductance can then be easily extracted
and depends on the number of contacts \( N_{R} \) and \( N_{L} \);
in particular the conductance differs in general from \( G_{0}=\frac{e^{2}}{h} \):
\begin{equation}
\label{chamon}
G(N_{L},N_{R})=\nu \frac{e^{2}}{h}\frac{\left[ 1-\left( \frac{\nu -1}{\nu +1}\right) ^{N_{L}}\right] \left[ 1-\left( \frac{\nu -1}{\nu +1}\right) ^{N_{R}}\right] }{\left[ 1-\left( \frac{\nu -1}{\nu +1}\right) ^{N_{R}+N_{L}}\right] }.
\end{equation}
\subsection{A toy model.}
We turn now to our model: it consists in a chiral Luttinger liquid
with Luttinger parameter \( \nu =1/(2n+1) \) in a circle geometry
with \( N \) terminals (Figure 3).  
That problem is easily solved by observing that it is a generalization
of Chamon and Fradkin model found by allowing each of \( N \) point
contacts to have distinct potentials. The \( n-th \) terminal has
potential \( V_{n} \), and due to the current \( i_{n} \) which
tunnels through it the chemical potential of the chiral LL is raised
from \( \mu _{n} \) to \( \mu _{n+1} \). The associated tunnel lagrangian
\( L^{n}_{tunnel} \) is exactly similar to \( L_{tunnel} \) given
above in eq.(\ref{l}), except for with the location of the tunneling
center. 

Let us focus on the \( n-th \) terminal; using equations (\ref{i_tunnel},\ref{i_tunnel2});
the current \( i_{n} \) is:\begin{eqnarray}
i_{n} & = & \frac{2\nu }{\nu +1}\frac{e}{h}\left( eV_{n}-\mu _{n}\right) ,\label{i4} \\
 & = & \nu \frac{e}{h}\left( \mu _{n+1}-\mu _{n}\right) .\label{i5} 
\end{eqnarray}
This yields the relation:\begin{equation}
\label{mu}
\mu _{n+1}=\alpha \mu _{n}+(1-\alpha )eV_{n}
\end{equation}
where 
$\alpha =\frac{\nu -1}{\nu +1}$.
Since we work on a circle with \( N \) terminals:
$\mu _{N+1}=\mu _{1}$;
we then deduce that each potential \( \mu _{n} \) can be completely expressed
in terms of the potentials of all the terminals:
\begin{equation}
\label{mu5}
\mu _{n}=\frac{(1-\alpha )\alpha ^{n-1}}{1-\alpha ^{N}}\left[ \sum ^{n-1}_{p=1}\frac{eV_{p}}{\alpha ^{p}}+\alpha ^{N}\sum ^{N}_{p=n}\frac{eV_{p}}{\alpha ^{p}}\right] .
\end{equation}
According to eq. (\ref{i5}), the current injected by the \( n-th \)
terminal $i_{n}=G_{nn}V_n +\sum_{p \ne n}G_{np}V_p$
yields the conductance matrix:
\begin{eqnarray}
G_{nn}= & \frac{e^{2}}{h}\frac{(1+\alpha )(1-\alpha ^{N-1})}{1-\alpha ^{N}} & \label{matrix1} \\
G_{np}= & -\frac{e^{2}}{h}\frac{(1-\alpha ^{2})\alpha ^{d(n,p)-1}}{1-\alpha ^{N}}, & n\neq p\label{matrix2} 
\end{eqnarray}
where $d(n,p)=n-p$ for $n>p$, or $d(n,p)=N+n-p$ for  $n<p$.

Several points are quite noteworthy:

(1) The boundary conditions described in \textbf{section 2} are implemented
exactly by eq.(\ref{mu5}) with the coefficients: 
\begin{eqnarray}
a_{n}^{p}=&\frac{(1-\alpha )\alpha ^{d(n,p)-1}}{1-\alpha ^{N}},& n \neq p \\
a_{n}^{n}=&\frac{(1-\alpha )\alpha ^{N-1}}{1-\alpha ^{N}},&
\end{eqnarray}

(2) In the scattering approach the conductance matrix is directly
related to probabilities of reflection and transmission through:
$G_{nn}=\frac{e^{2}}{h} \left( 1-R_{n}\right)$ and  $G_{np}=-\frac{e^{2}}{h} T_{np}$. Therefore \( G_{nn} \) is a positive number while \( G_{np} \) is
negative. However in our model \( G_{np} \) can be {\em positive} depending
on the value of \( d(n,p) \). 
More precisely \( G_{n,p} \) and \( G_{n,p+1} \) have alternating
signs: this means that even if electrodes \( p \) and \( p+1 \)
have the same potential currents issued from them are flowing in opposite
directions to electrode \( n \). This would have been impossible
for non-interacting electrons.

(3) Current conservation and gauge invariance are implemented since:
$\sum _{p}G_{pn}=0=\sum _{p}G_{np}$.

(4) The quantity \( \sum _{n,p}V_{n}V_{p}G_{np} \) is
always positive, which ensures dissipation of energy. 

Proof: It is equivalent to show that \( \sum _{n,p}V_{n}V_{p}\frac{1}{2}(G_{np}+G_{pn})\geq 0 \).
It is enough for that purpose to show that the eigenvalues of the
matrix \( \frac{1}{2}(G_{np}+G_{pn}) \) are all positive. But \( \frac{1}{2}(G_{np}+G_{pn}) \)
is a circulant matrix, i.e. a square matrix whose rows are obtained
by displacing the matrix elements of the first row by one column.
For a circulant matrix whose first row is \( (a_{1},..,a_{N}) \)
the \( k=1,...,N \) eigenvalues are equal to \( P(r_{k}) \) where
\( r_{k}=\exp \frac{i2\pi k}{N} \)is one of the \( N \)th roots
of unity and the polynomial \( P(X)=\sum _{p=0}^{N-1}a_{p+1}X^{p} \).
For the matrix \( \frac{1}{2}(G_{np}+G_{pn}) \) the first row \( (a_{1},..,a_{N}) \)
is given by:
$$\begin{array}{l}
a_{1}=G_{11}=\frac{e^{2}}{h}\frac{(1+\alpha )(1-\alpha ^{N-1})}{1-\alpha ^{N}},\\
a_{i}=\frac{1}{2}(G_{1i}+G_{i1})=-\frac{e^{2}}{h}\frac{(1-\alpha ^{2})}{1-\alpha ^{N}}\left( \frac{\alpha ^{i-2}+\alpha ^{N-i}}{2}\right) , i>1. 
\end{array}
$$
Therefore the eigenvalues are given by:\[
P(r_{k})=\frac{(1+\alpha )^{2}(1-\cos \frac{2\pi k}{N})}{(1+\alpha ^{2}-2\alpha \cos \frac{2\pi k}{N})}\geq 0.\]
 QED.

One eigenvalue vanishes (\( P(r_{0})=0 \)); it corresponds to the
eigenvector \( (V_{1},..,V_{N})=(1,..,1) \) which implements gauge
invariance (since the origin of voltages is arbitrary no current can
flow if all voltages are equal).

(5) It is easy to check that Onsager-Casimir reciprocity relations
are obeyed in this model: $G_{np}(\phi =\nu \phi _{0})=G_{pn}(\phi =-\nu \phi _{0})$. Since a magnetic field is present, under time-reversal one must reverse
its sign, which implies that \( \alpha  \) is changed into \( 1/\alpha  \).
Onsager-Casimir reciprocity then follows immediately from
eq. (\ref{matrix1}-\ref{matrix2}).)

(6) An interesting test-case of our model would be an experimental
setup with a FQHE disk for which the number of terminals can be changed
easily; what we envision is at first an experimental configuration
with of course few quantum point contacts (at least three in order
to observe the alternation of signs of non-diagonal elements of the
conductance matrix), which are tuned through a gate voltage, so that
the point contacts may be added or removed at will.

\section{Boundary conditions for a LL in a two-terminal geometry. }

\subsection{Boundary conditions}

As an illustration of our formalism we consider the simplest case
of a two-terminal geometry with a source and a drain at voltages \( V_S \)
and \( V_D \) (see Figure 4).

We take as a boundary condition the relation:

\begin{equation}
\label{bcsym}
\left( \begin{array}{c}
\mu _{+}\\
\mu _{-}
\end{array}\right) =A\left( \begin{array}{c}
eV_S\\
eV_D
\end{array}\right) 
\end{equation}
where the matrix \( \mathbf{A} \) is:\[
A=\left( \begin{array}{cc}
a^{S} & a^{D}\\
b^{S} & b^{D}
\end{array}\right) \]
 and where conservation of the number of particles imposes \( a^{S}+a^{D}=1=b^{S}+b^{D} \).
The parameter space is therefore two-dimensional. The conductance
matrix therefore is:\begin{eqnarray}
\left( \begin{array}{cc}
G_{SS} & G_{SD}\\
G_{DS} & G_{DD}
\end{array}\right)  
& =K\frac{e^{2}}{h}(a^{S}-b^{S})\left( \begin{array}{rr}
1 & -1\\
-1 & 1
\end{array}\right) .\nonumber \label{g} 
\end{eqnarray}
Therefore the two-terminal conductance \( G_{2t}=\frac{I}{V_S-V_D}=\frac{i_S}{V_S-V_D} \)
is:
\begin{equation}
G_{2t}=G_{SS}=K\frac{e^{2}}{h}(a^{S}-b^{S})=G_{K}\det A
\end{equation}
where the intrinsic conductance of the LL is \( G_{K}=\frac{Ie}{\mu _{+}-\mu _{-}}=K\frac{e^{2}}{h} \) (see \textbf{section 2}). 
Since the two-terminal conductance depends
on the difference \( (a^{S}-b^{S})=\det A \) distinct boundary conditions
or distinct coupling between the reservoirs and the LL can lead to
the same two-terminal conductance value. In contrast a given set of
the boundary conditions specifies unambiguously the conductance value.
If the two terminals are coupled in a symmetric manner to the LL one
has a further condition: \( a^{S}=b^{D} \). The parameter space is
then one-dimensional and \( G=K\frac{e^{2}}{h}(2a^{S}-1) \).

\subsection{Contact resistances.}

The LL has mean chemical potential \( \overline{\mu }=\frac{\mu _{+}+\mu _{-}}{2} \);
but the reservoirs have potentials \( eV_S \) and \( eV_D \).
Therefore there is a discontinuity between the chemical potentials
of the reservoirs and the LL. In the standard Landauer-Buttiker picture
of the contact resistance, the latter results precisely from such
a discontinuity at the boundaries of each reservoir\cite{key-25}:
on a length equal to the inelastic scattering length of each reservoir
collisions bring back the energy of each particle coming from the
metal to that of the reservoir. In our case, it follows immediately
from the boundary conditions that:
\begin{eqnarray}
eV_S-\overline{\mu }= & \frac{2-a^{S}-b^{S}}{2}\left( eV_S-eV_D\right)  & =\frac{h}{2Ke}\frac{2-a^{S}-b^{S}}{(a^{S}-b^{S})}I\label{mubc1} \\
\overline{\mu }-eV_D= & \frac{a^{S}+b^{S}}{2}\left( eV_S-eV_D\right)  & =\frac{h}{2Ke}\frac{a^{S}+b^{S}}{(a^{S}-b^{S})}I\nonumber 
\end{eqnarray}
which shows that there are two contact resistances:\begin{eqnarray*}
R_{S}= & \frac{R_{K}}{2}\frac{2-a^{S}-b^{S}}{(a^{S}-b^{S})} & \\
R_{D}= & \frac{R_{K}}{2}\frac{(a^{S}+b^{S})}{(a^{S}-b^{S})} & 
\end{eqnarray*}
where the intrinsic resistance of the LL is simply: \( R_{K}=1/G_{K} \).
These expressions also show that the two terminal conductance is obtained
from a series addition law of the two contact resistances: \( R_S+R_D=\frac{h}{Ke^{2}}\frac{1}{(a^{S}-b^{S})}=R_{2t}(=\frac{1}{G_{2t}}) \).
This implies that our boundary conditions incorporate an assumption
of \emph{incoherence between the contacts}. The above two equalities
eq.(\ref{mubc1}) are completely equivalent to the boundary conditions.
The two degrees of freedom in the boundary conditions simply reflect
the fact that there are two contact resistances. In the two terminal
geometry we may therefore rewrite the boundary conditions matrix \( A \)
in terms of the contact resistances:
\begin{equation}
\label{bcnew}
A=\frac{1}{(R_S+R_D)}\left( \begin{array}{rr}
R^{c}_{K}+R_D & \ -R^{c}_{K}+R_S\\
-R^{c}_{K}+R_D & \ R^{c}_{K}+R_S
\end{array}\right) ,
\end{equation}
 where we have defined an intrinsic contact resistance as: \( R^{c}_{K}=\frac{1}{2G_{K}}=\frac{1}{2KG_{0}} \)
.

Rewriting eq.(\ref{mubc1}) in terms of the chemical potentials leads
to an expression equivalent to the boundary condition expressed by
eq.(\ref{bcnew}):\begin{eqnarray}
eV_S-\overline{\mu } & =R_SIe & =R_SG_{2t}\left( eV_S-eV_D\right) \label{bcres} \\
\overline{\mu }-eV_D & =R_DIe & =R_DG_{2t}\left( eV_S-eV_D\right) \nonumber 
\end{eqnarray}

\subsection{Chamon and Fradkin model.}

In the Chamon-Fradkin model a Hall bar has two terminals on its left and on its
right \cite{key-34}. There are \( N_{R} \) (resp. \( N_{L} \))
point contacts at the right and left terminals. On the upper and lower
edges there are chiral Luttinger liquids flowing in opposite directions.
However \emph{the sum of the chiral hamiltonians for each chiral edge
is exactly identical to that of a non-chiral LL with parameter \( K=\nu  \).}
Through our phenomenological formalism, this allows us to describe
Chamon and Fradkin microscopic model of a chiral LL as a non-chiral
LL but with peculiar boundary conditions. If we make the reasonable
assumption that the contact resistances \( R_{S} \) and \( R_{D} \)
depend on \( N_{L} \) and \( N_{R} \) respectively (and not on both
\( N_{L} \) and \( N_{R} \)), there is a single boundary condition
corresponding to Chamon and Fradkin model. Given the two-terminal
conductance in eq.(\ref{chamon}) we find the boundary conditions
in eq.(\ref{bcnew}) with: \begin{equation}
\frac{1}{2R_{D}}=\nu \frac{e^{2}}{h}\frac{\left( 1-\left( \frac{\nu -1}{\nu +1}\right) ^{N_{R}}\right) ^{2}}{\left( 1-\left( \frac{\nu -1}{\nu +1}\right) ^{2N_{R}}\right) }
\end{equation}
\begin{equation}
\frac{1}{2R_{S}}=\nu \frac{e^{2}}{h}\frac{\left( 1-\left( \frac{\nu -1}{\nu +1}\right) ^{N_{L}}\right) ^{2}}{\left( 1-\left( \frac{\nu -1}{\nu +1}\right) ^{2N_{L}}\right) }
\end{equation}
This proves that for a non-chiral LL connected to two reservoirs it
is perfectly possible theoretically to have non-trivial contact resistances
(different from \( \frac{1}{2G_{0}} \)). The contact resistance  \( \frac{1}{2G_{0}} \) is retrieved for $N_R=N_L=1$.

\subsection{Comparison with earlier theories. }

Let us now discuss earlier theories of the two-terminal conductance
and show that the differences between their predicted values of \( G_{2t} \)
can be completely understood within our boundary condition formalism.
The differences stem from the hypotheses these theories make, implying
different boundary conditions and therefore different universality
classes for the joint system LL+electrodes. 

\subsubsection{Boundary conditions corresponding to \( G_{2t}=Ke^{2}/h \):}
Initially the conductance of the LL was thought to be \( G_{2t}=Ke^{2}/h \)
following the response function calculation \cite{3}.
Such calculation implicitly assumes an inversion symmetry between
source and drain. Comparing this conductance value with our computations
shows that \( R_S=R_D=R^{c}_{K} \). Going back to the
boundary conditions equations yields the universality class: \begin{equation}
\label{bcfq}
\left( \begin{array}{c}
\mu _{+}\\
\mu _{-}
\end{array}\right) =\left( \begin{array}{c}
eV_S\\
eV_D
\end{array}\right) .
\end{equation}
This allows a physical interpretation of the two-terminal conductance
\( G_{2t}=Ke^{2}/h \) for symetric electrodes : the conductance will
be equal to the Luttinger liquid parameter whenever there is an equilibrium
between a given reservoir and one of the two chiralities within the
LL. In other words the current injected by each electrode is completely
chiral. Such boundary conditions are realized in the FQHE: one finds
indeed a two-terminal conductance \( G_{2t}=Ke^{2}/h \) for a filling
fraction \( K \) which corresponds to a LL with parameter \( K \)
. It is also easy to show that eq.(\ref{bcfq}) is realized in the
FQHE for chiral edges. Indeed the chiral currents \( i_{\pm }=K\frac{e}{h}\mu _{\pm } \)
can be shown by using linear response to be also equal to \( i_{\pm }=K\frac{e}{h}\mu _{S/D} \)
which then implies immediately eq.(\ref{bcfq}). Such a conductance
can also be recovered for a non-chiral LL by using a Kubo formula
where it is assumed that an external field \( E_{0} \) applied on
a length \( L \) creates a voltage drop \( -E_{0}L=eV_{SD}=eV_S-eV_D=\mu _{+}-\mu _{-} \).
In our formalism the specification of the external electrical field
as \( E_{0}=-\frac{\mu _{+}-\mu _{-}}{L} \) is then understood as
implying an equilibrium between the reservoirs and a chirality of
the LL. But more generally this needn't be the case and a more general
linear relation between \( E_{0} \) and \( \mu _{+}-\mu _{-} \)might
hold, leading to another value of the two-terminal conductance.

As is clear from our formalism in absence of symetry between source
and drain the two terminal conductance value \( G_{2t}=Ke^{2}/h \)
can be obtained from any boundary conditions such that \( R_S+R_D=2R^{c}_{K} \). 

\subsubsection{Boundary conditions corresponding to \( G_{2t}=e^{2}/h \):} 
As we have already mentionned in the introduction many
other theoretical approaches predict a non-renormalized conductance
value \( G_{2t}=G_{0}=e^{2}/h \) per channel (these calculations
also implicitly assume a mirror symetry between source and drain so
that \( R_S=R_D \)). This shows that \( R_S=R_D=\frac{1}{2G_{0}} \)
so that the corresponding boundary conditions can be written in term
of the matrix \( Q[x] \) defined in eq.(\ref{q}) of \textbf{section
2}:\begin{equation}
\left( \begin{array}{c}
\mu _{+}\\
\mu _{-}
\end{array}\right) =Q[K^{-1}]\left( \begin{array}{c}
eV_S\\
eV_D
\end{array}\right) .
\end{equation}
or\begin{equation}
\label{cqfd}
\left( \begin{array}{c}
eV_S\\
eV_D
\end{array}\right) =Q[K]\left( \begin{array}{c}
\mu _{+}\\
\mu _{-}
\end{array}\right) .
\end{equation}
Besides for the inhomogeneous LL model (a LL for which the LL parameter
\( K(x) \) varies with position, \( K(0<x<L)=K \) and \( K(x)=1 \)
otherwise) by using continuity equations for the phase fields across
the boundaries it can be explicitly shown that\cite{key-36}:\begin{equation}
\label{cqfd1}
\left( \begin{array}{c}
eV_S\\
eV_D
\end{array}\right) =\left( \begin{array}{c}
\mu ^{0}_{+}\\
\mu ^{0}_{-}
\end{array}\right) ,
\end{equation}
where \( \mu ^{0}_{\pm }=\frac{\partial H}{\partial N^{0}_{\pm }} \)
are the chiral potentials for non-interacting electrons, that is if
\( K=1 \) (\( N^{0}_{\pm } \) is the number of fermions at the right
or left Fermi points). But this can be reexpressed in terms of the
chiral chemical potentials by noting that the chiral densities are
related to the left or right moving fermions densities by eq.(\ref{rho2}),
which results in \begin{equation}
\label{cqfd2}
\left( \begin{array}{c}
\mu ^{0}_{+}\\
\mu ^{0}_{-}
\end{array}\right) =Q[K]\left( \begin{array}{c}
\mu _{+}\\
\mu _{-}
\end{array}\right) .
\end{equation}
Using the eq.(\ref{cqfd1},\ref{cqfd2}) leads immediatlely to our
boundary conditions eq.(\ref{cqfd}). 

For completeness, let us mention several papers which have already
developped a narrower boundary conditions point of view to describe
linear transport of LL in a two-terminal geometry.  Fröhlich et al.
implicitly considered the boundary conditions \( \left( \begin{array}{c}
eV_S\\
eV_D
\end{array}\right) =\left( \begin{array}{c}
\mu ^{0}_{+}\\
\mu ^{0}_{-}
\end{array}\right)  \) or \( \left( \begin{array}{c}
eV_S\\
eV_D
\end{array}\right) =\left( \begin{array}{c}
\mu _{+}\\
\mu _{-}
\end{array}\right)  \) for a LL and for a chiral LL for the FQHE in reference \cite{key-8};
these are subcases of our own formalism. Egger et al. then discussed
so-called radiative boundary conditions for the LL \cite{key-9},
and B\"uttiker et al. derived a similar boundary condition later \cite{key-10}.
Safi showed finally in \cite{key-11} that the boundary conditions
of \cite{key-9,key-10} are all equivalent to \( \left( \begin{array}{c}
eV_S\\
eV_D
\end{array}\right) =\left( \begin{array}{c}
\mu ^{0}_{+}\\
\mu ^{0}_{-}
\end{array}\right)  \) . 
Therefore all these approaches boil down within our formalism to
choosing one particular boundary condition. 

\subsubsection{Other boundary conditions?} 
Within our formalism, earlier
theories correspond to two particular boundary conditions leading
either to \( G=Ke^{2}/h \) or to \( G=e^{2}/h \). But as discussed
in the introduction experimental evidence on carbon nanotubes and
quantum wires yield conductance plateaux at values which differ from
\( G=G_{0} \) per channel \cite{key-15,key-16,key-19,key-21}.

The case of quantum wires fabricated by the cleaved edge overgrowth
technique is quite noteworthy: on the one hand evidence points towards
a LL physics \cite{key-16}; on the other hand Landauer scattering
approach is used to interpret conductance measurements in \cite{key-19}.
Yet Landauer-B\"uttiker formalism is invalid in that context of strongly
correlated electrons! Our formalism resolves the tension because it
includes a theory of linear transport which is independent of Landauer
scattering approach, while being applicable to the LL. 

Whether more general boundary conditions than those implicit in earlier
theories are realized must be settled by experiments. But we observe
firstly that the Chamon and Fradkin model for two chiral edges can be
interpreted as a non-chiral LL connected (albeit in a very peculiar
manner) to two large reservoirs; this model then yields a conductance
\( G\neq G_{0} \): this implies that at least theoretically there
is no grounds for a no-go theorem preventing two-terminal conductance
plateaux at values distinct from the quantum of conductance. Secondly
it is noteworthy that in the inhomogeneous LL model which falls into
the class of boundary condition \( \left( \begin{array}{c}
eV_S\\
eV_D
\end{array}\right) =\left( \begin{array}{c}
\mu ^{0}_{+}\\
\mu ^{0}_{-}
\end{array}\right)  \) \emph{injection of current from the reservoirs to the LL is done
through a single point contact}. But experimentally both quantum wires
and carbon nanotubes have a large area in contact with the reservoirs:
whether one can safely assume that there is a single tunneling point
contact is therefore extremely doubtful. It it then perfectly conceivable
that other boundary conditions than \( \left( \begin{array}{c}
eV_S\\
eV_D
\end{array}\right) =\left( \begin{array}{c}
\mu ^{0}_{+}\\
\mu ^{0}_{-}
\end{array}\right)  \) may be valid, leading therefore to a conductance \( G\neq G_{0} \).
The quantum wire or individual single wall carbon nanotubes experiments
finding conductances unquantized at \( G_{0} \) may therefore be
explainable using interacting electrons within our formalism.

The variety of boundary conditions reflects simply the nature of the
equilibrium achieved between the reservoirs and the LL: the chemical
potentials of the charge carriers within the LL (i.e. the chiral chemical
potentials) need not be identical with those of the charge reservoirs.
Only for symetric coupling of the electrodes and if \( a^{S}=1 \)
does one find that \( eV_S=\mu _{+} \) and \( eV_D=\mu _{-} \):
Kane and Fisher calculations fall into that class. That boundary condition
is natural for the FQHE because it is sensible for the chemical potentials
of the reservoirs to be in equilibrium with those of the charge carriers,
since the contact region between the FQHE condensate and the reservoirs
is large. However as shown microscopically by Chamon et al.\cite{key-34},
if the contact with the reservoirs is not perfect (e.g. a granularity
limits the number of tunneling points), such an equilibrium may not
be achieved.

\section{Two-terminal geometry with an impurity. }

\subsection{New boundary conditions.}

We now insert a weak local impurity in the wire: \begin{equation}
V=u\delta (x)(\Psi _{R}^{+}\Psi _{L}+h.c.).
\end{equation}
A current will therefore be backscattered and the potentials need not
be identical across the impurity. We use eq.(\ref{bcres}) to write
boundary conditions in the presence of an impurity:\begin{eqnarray}
eV_S-\overline{\mu }_{L} & =R_SIe & \label{bcimp} \\
\overline{\mu }_{R}-eV_D & =R_DIe & \nonumber 
\end{eqnarray}
where the index \( R/L \) refers to \( \mu ^{\pm }_{R/L} \) are
the chiral chemical potentials to the right or the left of the impurity
and \( \overline{\mu } \) is the average between the
chiral chemical potentials. To have conditions on the sole chemical
potentials it suffices then to remark that \( I=\frac{K\: e}{h}\: \left( \mu ^{+}_{L}-\mu _{L}^{-}\right) =\frac{K\: e}{h}\: \left( \mu ^{+}_{R}-\mu _{R}^{-}\right)  \),
so that:\begin{eqnarray}
eV_S-\overline{\mu }_{L} & =R_SG_{K}\: \left( \mu ^{+}_{L}-\mu _{L}^{-}\right)  & \label{bcimp1} \\
\overline{\mu }_{R}-eV_D & =R_DG_{K}\: \left( \mu ^{+}_{R}-\mu _{R}^{-}\right)  & \nonumber 
\end{eqnarray}
In spite of these linear relations the chiral chemical potentials
need not depend linearly on the external voltages (there are four
chemical potentials for two linear relations). There is also a non-linear
contribution due to the backscattering at the impurity.

\subsection{What is the backscattering current? }

In the presence of the impurity, some of the current is backscattered.
It is usually assumed that the backscattering current is simply the
difference between the current in the absence of impurity and the
current in the presence of the impurity. This is only correct for
non-interacting systems, but for a LL this will depend on both the
conductance \( G_{2t} \) and \( K \). We must go back to the definition
of the backscattering current as the velocity times the density difference
of right-movers on the left and on the right of the impurity:\begin{equation}
i_{B}=u \ e \left( \rho _{L}^{+}-\rho _{R}^{+}\right) .
\end{equation}
 This is also equal by charge conservation to: \( u \ e \left( \rho _{L}^{-}-\rho _{R}^{-}\right)  \)
with obvious notations. We can also relate \( i_{B} \) to the chiral
chemical potentials by using the fact that \( u \ \rho _{L}^{+}=\frac{K}{h}\ \mu _{L}^{+} \)(with similar relations for the \( - \) 
chirality and for the potentials to the right of the impurity). 
Therefore:
\begin{equation}
i_{B}=\frac{K \ e}{h} \left( \mu ^{+}_{L}-\mu _{R}^{+}\right) =\frac{K \ e}{h} \left( \mu ^{-}_{L}-\mu _{R}^{-}\right) .
\end{equation}
Using the new boundary conditions (\ref{bcimp1}):
\begin{eqnarray*}
eV_S-eV_D & = & \overline{\mu }_{R}+R_DIe+\overline{\mu }_{L}+R_SIe\\
 & = & R_{K}i_{B}\, e+(R_S+R_D)Ie
\end{eqnarray*}
where \( R_{K}=h/Ke^{2} \). This can be recast as:
\begin{equation}
\begin{array}{ll}
I&=I_{0}-\frac{R_{K}}{R_S+R_D}i_{B} \label{ib2} \\
&\neq I_{0}-i_{B} 
\end{array}
\end{equation}
where \begin{equation}
\label{io}
I_{0}=\frac{eV_S-eV_D}{e\: (R_S+R_D)}
\end{equation}
is the current in the absence of an impurity and is therefore also
the saturation current, i.e. the maximal current which can be reached
when one goes to large voltages. The fact that \( i_{B}\neq I_{0}-I \)
contrary to the naive expectation stems from the contact resistances:
the difference between \( i_{B} \) and \( I-I_{0} \) is akin to
the difference between a two-teminal and a four-terminal measurement.
\( I_{0}-I \) takes into account the resistance at the contacts while
\( i_{B} \) is more intrinsic and measures the net current wich is
backscattered \emph{locally} at the impurity.

\section{Shot noise in a two-terminal geometry. }

What are the elementary excitations of the LL? The textbook answer
is that there are two kinds of excitations: (1) bosonic density fluctuations
(plasmons); (2) zero modes ladder operators which change the number
of particles at each Fermi point but have no dynamics \cite{3}. It
is seldom remarked that such a description of the excitations found
for the LL through the bosonization method is also valid for free
electrons. What this means is that for free electrons there are two
equivalent manners of describing the elementary excitations (corresponding
to two basis of eigenstates): (1) the usual manner, in terms of charged
quasiparticles (the electron and the hole); (2) and the one provided
by bosonization, which yields bosonic density fluctuations and ladder
operators. The two descriptions differ markedly in that the second
involves charged excitations which have no dispersion, while in the
first the charge dynamics is described by the usual quasiparticles.

For the LL it can be shown that exactly in the same manner there exists
a basis of charged quasiparticles. However instead of the Landau quasiparticle
one finds fractional elementary excitations, which may even carry
irrational charges. In particular the particle-hole continuum of Fermi
liquid theory is replaced by a quasiparticle-quasihole continuum of
excitations which are the analogs of Laughlin quasiparticles\cite{4}.
For the chiral LL (the edge states of the FQHE) they have been detected
through shot noise. \emph{In the case of the non-chiral LL a marked
difference is that such shot noise experiments would allow to detect
irrational charges} (the FQHE filling fraction \( \nu  \) which is
a rational number is replaced by the LL parameter \( K \)).

Present theories of shot noise can be roughly separated into two camps:
{\bf A} Kane, Fisher, Balents et al.\cite{key-26,key-31} predict a Fano factor equals to \( Ke \). This is commonly interpreted as the proof that excitations of charge \( Ke \) are responsible for the noise. This calculation however makes no explicit modelization of the reservoirs; {\bf B} Ponomarenko
et al., Egger et al.\cite{key-27} work with the inhomogeneous LL 
(two terminal geometry which models the reservoirs 
as 1D Fermi liquids on a half-line) and
find a Fano factor or excitations of charge equals to \( e \). 
We note that Blanter and B\"uttiker have argued
against this last result by noting that the shot noise should
not depend on the reservoirs since this is a measure of the charge
backscattered \emph{locally} by the impurity. We discuss now these
two sets of theories: {\bf A} we apply our boundary conditions formalism
to the shot noise theory of Kane and Fisher; {\bf B} for the inhomogeneous
LL we discuss the meaning of their result in the light of
the identities derived in the previous section.

\subsection{Kane-Fisher approach}

The shot noise through a weak impurity in a LL was first computed
by Kane and Fisher (for the edge states of the FQHE and before the
actual proof that there exists also Laughlin quasiparticles in the
LL) by using the Keldysh formalism applied to an effective lagrangian
found by integrating out the degrees of freedom away from an unique impurity
(\( \phi  \) is the standard LL phase field at the location of the
impurity)\cite{key-26}:
\begin{equation}
L=\frac{1}{2K}\sum _{n}\left| \omega _{n}\right| \left| \Phi (i\omega _{n})\right| ^{2}+\int d\tau \! v\cos (2\sqrt{\pi }\phi (\tau )).
\end{equation}

Although initially intended for the edge states of the FQHE the calculation
is also valid for the non-chiral LL. Kane and Fisher find that the
current and the noise are given respectively by:
\begin{eqnarray}
I & = & K\frac{e^{2}}{h}\left( \frac{da}{dt}-V\right) ,\label{courantn} \\
S_{I} & = & Ke\left( K\frac{e^{2}}{h}\frac{da}{dt}-I\right) \label{bruit} 
\end{eqnarray}
where \( V=\left\langle \frac{h}{e}v\sin \left( 2\sqrt{\pi }(\phi +Ka\right) \right\rangle  \),
with a source term \( \int j.a \) added to the lagrangian. 
Kane and Fisher assumed that 
\begin{equation}
\label{da/dt}
e\frac{da}{dt}=e(V_S-V_D).
\end{equation}

In the absence of impurity \( V=0 \), eq.(\ref{courantn}) leads to 
\begin{equation}
\label{i0}
I_{0}=K\frac{e^{2}}{h}\frac{da}{dt},
\end{equation}
on the other hand according
to eq.(\ref{courant}), \( I_{0}=K\frac{e}{h}\left( \mu _{+}-\mu _{-}\right)  \), together with eq. (\ref{da/dt}) this then implies that \( e(V_S-V_D)=\left( \mu _{+}-\mu _{-}\right)  \). Therefore within Kane-Fisher approach and assuming eq. (\ref{da/dt}) the ``two terminal'' conductance value is \( G_{2t}=Ke^{2}/h \)  in the absence of an impurity.
Moreover as discussed in \textbf{section
4.4} (assuming symmetric coupling to source and drain) this calculation
falls into the class of boundary conditions which correspond to \( \left( \begin{array}{c}
\mu _{+}\\
\mu _{-}
\end{array}\right) =\left( \begin{array}{c}
eV_S\\
eV_D
\end{array}\right)  \), i.e. equilibrium of the reservoirs chemical potentials with those of the LL. 
In order to obtain values of the conductance different from \( Ke^{2}/h \), it
is sufficient to change the previous assumption eq. (\ref{da/dt}): 
other classes of boundary conditions are found simply 
by \emph{assuming that the response
of the LL is totally driven by the values of the chiral chemical potentials}
(\emph{the chemical potentials of the charge carriers of the LL) in
the absence of an impurity,} and not by the reservoirs potentials
(since there is no reason why they should be equal). We therefore
modify eq.(\ref{da/dt}) into: 
\begin{eqnarray}
e\frac{da}{dt} & = & \mu _{+}-\mu _{-} \\
 & = & \frac{R_{K}}{R_S+R_D}e(V_S-V_D)\label{hyp}
\end{eqnarray}
where in the second line the boundary conditions (\ref{bcnew}) in
the absence of impurity have been used. 
Eq. (\ref{hyp}) together with eq. (\ref{i0}) then lead
to the value of the ``two-terminal'' conductance
 that we obtained with our approach in {\bf section 4} 
for the same boundary conditions, namely \( G_{2t}=1/(R_S+R_D)\).


So far we have shown that by assuming the source term 
definition Eq. (\ref{hyp}) instead of Eq. (\ref{da/dt}) 
it is possible to adapt Kane-Fisher calculations 
to reproduce the various boundary conditions in the absence of impurity. 

We can now reconsider Kane and Fisher's calculations for the shot noise, 
i.e. in the presence of an impurity in the bulk of the LL. 
More precisely we express now the
shot noise as a function of either \( I_{0}-I \) 
(the deviation to the saturation current) 
or as a function of the backscattering current \( i_{B} \),
for the various boundary conditions.
According to eq. (\ref{i0}) and eq. (\ref{bruit}) we always 
have \emph{independently of the boundary condition chosen}:
\begin{equation}
\label{bruit2}
S_{I}=Ke\! \left( I_{0}-I\right).
\end{equation}
Therefore, using eq. (\ref{ib2}) \( I=I_{0}-\frac{R_{K}}{R_S+R_D}i_{B}=I_{0}-\frac{G_{2t}}{G_{K}}i_B \) the last equality is also recast as:
\begin{equation}
\label{bruit3}
S_{I}=\frac{G_{2t}}{G_{0}}\; e\: i_{B},
\end{equation}
where $G_{2t}$ is the two terminal conductance 
that reflects the contact resistances in the absence of the impurity.
The shot noise 
Fano factor might therefore appear to depend on whether one refers
to the backscattering current \( i_{B} \) or to \( I_{0}-I \), the
deviation to the saturation current. But since \( S_{I} \) is the
fluctuation of the current \( I \), the physical shot noise charge
must be measured with respect to the current \( I \) and not with
respect to \( i_{B} \). The shot noise charge
is therefore \( \frac{_{S_{I}}}{I_{0}-I}=Ke \), independently of
the boundary condition realized in the system and is not equal to
\( \frac{S_{I}}{i_{B}} \).

At any rate what is directly measured is always \( I \) or \( I_{0} \):
\( i_{B} \) is only indirectly accessible through for instance eq.
(\ref{ib2}).

In summary, within the Kane and Fisher calculation by the Keldysh method,
it is therefore possible to have (i) an ohmic conductance distinct from
\( Ke^{2}/h \) and (ii) a shot noise charge equal to \( Ke \) 
independently on the value of the ohmic conductance.

\subsection{Inhomogeneous model approach}

The role of the reservoirs on the measure of the shot noise of a LL 
was examined in two papers \cite{key-27}, 
which make calculations on the inhomogeneous
LL model for which the boundary condition is \( \left( \begin{array}{c}
eV_S\\
eV_D
\end{array}\right) =Q[K]\left( \begin{array}{c}
\mu _{+}\\
\mu _{-}
\end{array}\right)  \) as discussed in section \textbf{4.4}. 
They both find that \( S_{I}=e\! (I_{0}-I) \)
with a conductance \( G_{2t}=G_{0} \) . It is interesting to remark that
the equation \( S_{I}=e\: (I_{0}-I) \) can be recasted in terms of
the backscattering current noise as:
\[S_{i_{B}}=Ke\: i_{B},\]
since \( I_{0}-I=i_{B}/K \) when \( G_{2t}=G_{0} \).
 

The result of Ponomarenko et al. and Egger et al. \cite{key-27}, 
acquires then the 
following interpretation: the charge \( Ke \) is not found in the
shot noise for the total current in the LL because it is really the
correlations of the backscattering current which should be measured.
Since the impurity backscatters charge \( Ke \) Laughlin quasiparticles,
the backscattering current correlations must contain the information
on the charge backscattered by the impurity.

This is in disagreement with Kane and Fisher theory even when this last
theory is modified by our boundary conditions formalism in order to reproduce 
the situation $G_{2t}=G_0$. We are unable as yet to explain the discrepancy
between the two approaches. 


Lastly, we note that taking the relation
\( S_{i_{B}}=Ke\: i_{B} \) as a starting point, 
this then according to our formalism leads inevitably 
to \( S_{I}=\frac{G_{2t}}{G_{0}}\; e\: (I_{0}-I) \).
An experimental test of this last suggestion would require an independant
measurement of $K$ and $i_B$. In FQHE such independant measurement 
is possible because the two chiralities of the effective LL are 
physically well separated and a direct measure 
of $K$ is then possible through the Hall conductance. 
In contrast to the FQHE, \( i_{B} \) 
is not experimentally measureable in Carbon nanotubes: 
this then means in turn that even though 
\( S_{i_{B}}=Ke\: i_{B} \) 
the charge of Laughlin quasiparticles in a non-chiral LL is not directly
measurable through shot noise in a two-terminal geometry.

Shot noise experiments will hopefully settle the issue. In this respect
some experiments on carbon nanotubes are in progress \cite{key-29}.

\section{Conclusions.}

We proposed in this paper a new formalism which modelizes the joint
system LL+electrodes as a single LL with no electrodes but subjected
to boundary conditions on its chiral chemical potentials. We were
able to show in a solvable toy-model that such boundary conditions
can indeed be derived explicitly. That model is quite remarkable because
the conductance matrix of the LL in contact with an arbitrary number
of terminals can be computed; it is found that the probabilistic scattering
approach fails: it would lead to negative probabilities for the transmission
of electrons. The obvious advantage of our formalism is that it avoids
discussion of the detailed microscopics of a system, but yields a
classification of the joint system LL+electrodes and then makes precise
predictions for the transport. In particular the Landauer-B\"uttiker
view of the contact resistance as resulting from a mismatch between
chemical potentials is recovered; if the charge backscattered by an
impurity in a LL is \( \frac{S_{i_{B}}}{i_{B}} \) we find that the
shot noise of the total current does not allow a measure of the fractional
charge \( Ke \) of Laughlin particles in a LL. 

It is easy to generalize our formalism to the case of several channels:
(i) several conducting channels as in carbon nanotubes; (ii) spin
transport: this arises with ferromagnetic reservoirs; (iii) application
of a magnetic field on the LL, which breaks the spin-charge separation.

The authors wish to thank M. Gabay for useful discussions.

\begin{figure}[h]
\epsfig{figure=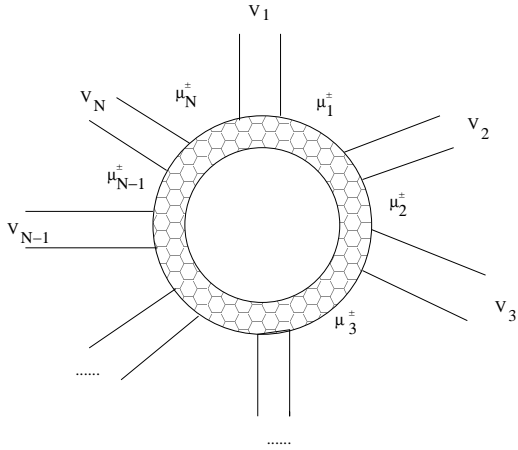,width=7.cm}
\caption{LL connected to many electrodes in a loop geometry. $V_n$ is the potential of electrode $n$. $\mu_n ^{\pm}$ are the chiral chemical potential on the left of electrode $n$. Due to current injection at each terminal the chiral chemical potentials may change their values across them: accordingly \( \mu _{n}^{\pm} \) on the left of terminal \( n \) is changed to \( \mu _{n+1}^{\pm} \) on the right of the same terminal.}
\end{figure}

\begin{figure}[h]
\epsfig{figure=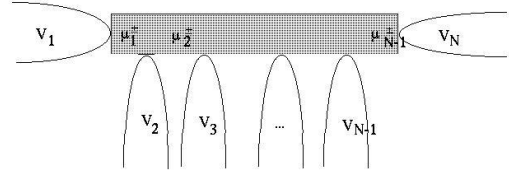,width=7.cm}
\caption{Same as Fig. 1 but for LL connected to many electrodes 
in a rod geometry.}
\end{figure}

\begin{figure}[h]
\epsfig{figure=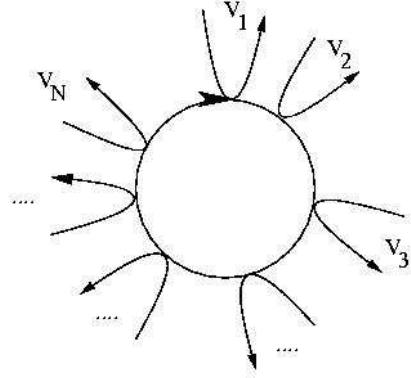,width=7.cm}
\caption{Model of a chiral LL connected to many chiral electrodes 
in a loop geometry.}
\end{figure}

\vspace{1.0cm}
\begin{figure}[h]
%
\epsfig{figure=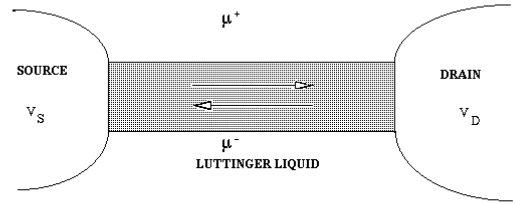,width=7.cm}
\caption{LL in a two terminals geometry.}
\end{figure}

\end{multicols}

\end{document}